%% file: ms.tex
\documentclass[usenatbib]{mn2e}
\usepackage{natbibmnfix, graphicx, times, amsmath, epsfig, amssymb}
\include{defns}

\pdfoutput=1

\begin{document}

\title[The impact of UVB feedback on reionization]{How does radiative feedback from a UV background impact reionization?}
\author[Sobacchi \& Mesinger]{Emanuele Sobacchi$^1$\thanks{email: emanuele.sobacchi@sns.it}, Andrei Mesinger$^1$  \\
$^1$Scuola Normale Superiore, Piazza dei Cavalieri 7, 56126 Pisa, Italy\\
}

\voffset-.6in

\maketitle

\begin{abstract}
An ionizing UV background (UVB) inhibits gas accretion and photo-evaporates gas from the shallow potential wells of small, dwarf galaxies.  During cosmological reionization, this effect can result in negative feedback: suppressing star-formation inside HII regions, thus impeding their continued growth.  It is difficult to model this process, given the enormous range of scales involved. 
We tackle this problem using a tiered approach: combining parameterized results from single-halo collapse simulations with large-scale models of reionization. In the resulting reionization models, the ionizing emissivity of galaxies depends on the {\it local} values of the reionization redshift and the UVB intensity.
We present a physically-motivated analytic expression for the average minimum mass of star-forming galaxies, $\bar{M}_{\rm min}$, which can be readily used in modeling galaxy formation.
We find that UVB feedback: (i) delays the end stages of reionization by $\Delta z \lsim 0.5$; (ii) results in a more uniform distribution of HII regions, peaked on smaller-scales (with large-scale ionization power suppressed by tens of percent); and (iii) suppresses the global photoionization rate per baryon by a factor of $\lsim2$ towards the end of reionization.
  However, the impact is modest, since the hydrodynamic response of the gas to the UVB occurs on a time-scale comparable to reionization.  In particular, the popular approach of modeling UVB feedback with an instantaneous transition in $M_{\rm min}$, dramatically overestimates its importance.  UVB feedback does not significantly affect reionization unless: (i) molecularly-cooled galaxies contribute significantly to reionization; or (ii) internal feedback processes strongly couple with UVB feedback in the early Universe.  Since both are considered unlikely, we conclude that there is no significant self-regulation of reionization by UVB feedback.
\end{abstract}


\section{Introduction}
\label{sec:intro}

Early generations of luminous objects reionized the Universe within the first $\sim$billion years following the Big Bang.  This process is expected to be fairly extended and inhomogeneous.  As ionizing radiation spread out from early galaxies, it heated the intergalactic medium (IGM), affecting its cooling properties, and photo-evaporated gas from shallow potential wells.  In this picture, galaxies forming inside an already-ionized IGM would have a depleted gas reservoir (and by extension, fewer stars) compared with those forming inside the neutral IGM.  Therefore, the ionizing ultraviolet background (UVB) during reionization results in a negative feedback mechanism, hindering the further growth of ionized regions. This negative feedback could delay reionization, as well as result in a more uniform reionization morphology.  Therefore understanding UVB radiative feedback is crucial in developing models of reionization, as well as interpreting observations.

The main difficulty in self-consistently modeling UVB feedback lies in the enormous dynamical range required.  Ideally, we would need to fully resolve the internal structure of the dominant population of atomically-cooled halos (with virial radii $\sim 1\text{ proper kpc}$; \citealt{BL01}) in cosmological simulations which resolve ionizing structure on $\gsim 100\text{ comoving Mpc}$ scales (e.g. \citealt{FHZ04}). Even state-of-the-art single galaxy simulations must employ some analytic prescriptions in estimating internal feedback (e.g. \citealt{HQM11}); therefore compromises must be made.
Generally one of two approaches are followed.
The first approach uses relatively small simulation boxes ($\sim$ few cMpc on a side), attempting to model the sub-grid interstellar medium (ISM) physics, through analytical prescriptions (e.g. \citealt{PS11, FDO11, WTNA12, JDK12, HS12}).
Here it is important to explore reasonable extremes of all sub-grid recipes, and ensure resolution convergence.  Even so, it is currently impossible to model the relevant scales of reionization.  The second approach simulates reionization on large-scales, but does not model the internal structure of galaxies; instead, simple analytic (and relatively ad-hoc) prescriptions are used to account for UVB feedback (e.g. \citealt{IMSP07, QLZD07}).

In this paper we follow the later approach.  However, rather than adopting ad-hoc prescriptions, we base our treatment of UVB feedback on results from parameter studies of 1D collapse simulations (\citealt{SM12a}, hereafter Paper I).  In Paper I, we obtained expressions for the baryon content of galaxies, residing in regions which were reionized at redshift $z_{\rm IN}$.  In this work, we include these expressions in semi-numerical simulations of inhomogeneous reionization.  We present large-scale reionization simulations which include a physically-motivated prescription for UVB feedback.  With these simulations, we study the importance of UVB feedback in regulating reionization.

This paper is organized as follows.  In \S \ref{sec:reion} we describe our UVB feedback prescription, and our semi-numerical reionization simulations. In \S \ref{sec:results} we present our results, comparing different inhomogeneous reionization models.
In \S \ref{sec:concl} we discuss our results and present our conclusions. Throughout we assume a flat $\Lambda\text{CDM}$ cosmology with parameters ($\Omega_{\rm m}$, $\Omega_{\Lambda}$, $\Omega_{\rm b}$, $h$, $\sigma_{\rm 8}$, $n$) = (0.27, 0.73, 0.046, 0.7, 0.82, 0.96), consistent with WMAP results \citep{WMAP11}.  Unless stated otherwise, we quote all quantities in comoving units.

\section{Modeling reionization with UVB feedback}
\label{sec:reion}

\subsection{Minimum halo mass to host galaxies}
\label{sec:Mmin}

In the simplest UVB-regulated scenario\footnote{We caution that we do not directly include internal feedback processes, which, depending on the ISM treatment, could be important in determining the star formation rate (SFR) of dwarf galaxies at high-redshifts (e.g. \citealt{PS09, FDO11, WL12}).
Although winds can push gas to the outskirts of galaxies, where it can be more easily photo-evaporated by a UVB, this amplification effect is modest at high-redshifts (at redshifts not far below $z_{\rm IN}$; e.g. \citealt{PS09, FDO11}).
Hence, during most of reionization, it is probably reasonable to model internal feedback independently from UVB feedback (e.g. by decreasing the ionizing efficiency parameter, $\xi$, below).  Since we are primarily concerned with the relative impact of UVB feedback, we do not expect internal feedback to alter our main conclusions.},
there are two fundamental halo mass scales regulating the gas reservoir available for star formation: (\textit{i}) $M_{\rm crit}$, the characteristic or critical mass below which baryons are photo-evaporated or cannot efficiently accrete onto their host halos; (\textit{ii}) $M_{\rm cool}$, the cooling threshold required for gas to efficiently cool, collapse and form stars inside the halo.
 Therefore, we can define the minimum mass of star-forming halos as:
\begin{equation}
\label{eq:M_min}
M_{\rm min}=\max\left[M_{\rm cool}, M_{\rm crit}\right] ~ .
\end{equation}
\noindent Here we mainly consider the advanced stages of reionization, where cooling through molecular hydrogen is expected to be relatively inefficient due to a disassociating background (e.g. \citealt{HRL97}). This corresponds to an atomic cooling threshold of a constant virial temperature, $T_{\rm vir}\sim10^4$ K, or analogously a redshift-dependent halo mass (e.g. \citealt{BL01}):
\begin{align}
\label{eq:M_cool}
M_{\rm cool} & = 10^{8}h^{-1} \left(\frac{\mu}{0.6}\right)^{-3/2} \left(\frac{\Omega_{\rm m}}{\Omega_{\rm m}^{\phantom{\text{ }}\rm z}} \frac{\Delta_{\rm c}}{18\pi^{2}}\right)^{-1/2} \times \nonumber \\ & \times \left(\frac{T_{\rm vir}}{1.98\times 10^{4}\text{ K}}\right)^{3/2} \left(\frac{1+z}{10}\right)^{-3/2} M_{\odot} \simeq \nonumber
\\  & \simeq 10^{8}\left(\frac{1+z}{10}\right)^{-3/2}M_{\odot}
\end{align}
where $\mu$ is the mean molecular weight, $\Omega_{\rm m}^{\phantom{\text{ }}\rm z}=\Omega_{\rm m}\left(1+z\right)^{3}/\left[\Omega_{\rm m}\left(1+z\right)^{3}+\Omega_{\Lambda}\right]$ and $\Delta_{\rm c}=18\pi^{2}+82d-39d^{2}$ with $d=\Omega_{\rm m}^{\phantom{\text{ }}\rm z}-1$.

Fitting to a broad parameter space of 1D collapse simulations, in Paper I we presented the following expression for $M_{\rm crit}$ (technically defined as the total halo mass at which the baryon fraction is 1/2 of the global value, $\Omega_{\rm b}/\Omega_{\rm m}$):
\begin{equation}
\label{eq:critical_mass_3}
M_{\rm crit}=M_{\rm 0}J_{\rm 21}^{\phantom{21}a}\left(\frac{1+z}{10}\right)^{b}\left[1-\left(\frac{1+z}{1+z_{\rm IN}}\right)^{c}\right]^{d}
\end{equation}
\noindent with best-fit parameters:
\begin{equation}
\label{eq:fit_const}
\left(M_{\rm 0}, a, b, c, d\right)=\left(2.8\times 10^{9}M_{\odot}, 0.17, -2.1, 2.0, 2.5\right)
\end{equation}
\noindent The critical mass depends on (\textit{i}) redshift $z$; (\textit{ii}) the redshift $z_{\rm IN}$ when the halo is exposed to a UVB; (\textit{iii}) the UVB intensity $J_{\rm 21}$:\footnote{In Paper I, we ignored self-shielding when computing the critical mass. The importance of self-shielding on such dwarf galaxies at high-redshifts is not well known, with some authors suggesting that it does not have a large impact on the relevant photo-evaporation time-scales (e.g. \citealt{SIR04, ISR05}). Nevertheless, it should be noted that neglecting self-shielding overestimates the impact of UVB feedback on reionization, making our main conclusions conservative.}
\begin{equation}
\label{J21}
J\left(\nu\right)=J_{\rm 21}\left(\nu/\nu_{\rm H}\right)^{-\alpha}\times 10^{-21}\text{ erg s$^{-1}$ Hz$^{-1}$ cm$^{-2}$ sr$^{-1}$}
\end{equation}
\noindent where $\nu_{\rm H}$ is the Lyman limit frequency and $\alpha=5$ corresponds to a stellar-driven UV spectrum (e.g. \citealt{TW96}).

\subsection{Semi-numerical reionization simulations}
\label{sec:code_descr}

\begin{figure*}
\vspace{+0\baselineskip}
{
\includegraphics[width=0.95\textwidth]{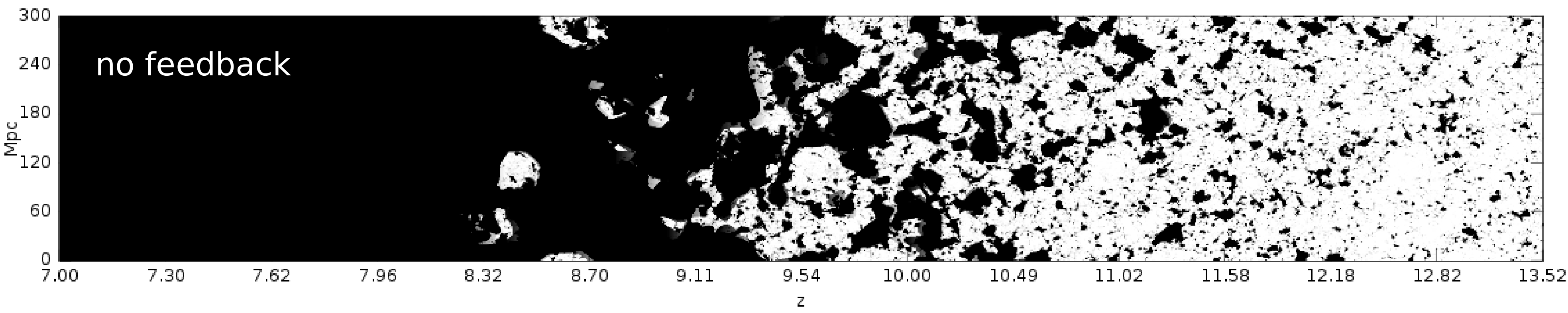}
\includegraphics[width=0.95\textwidth]{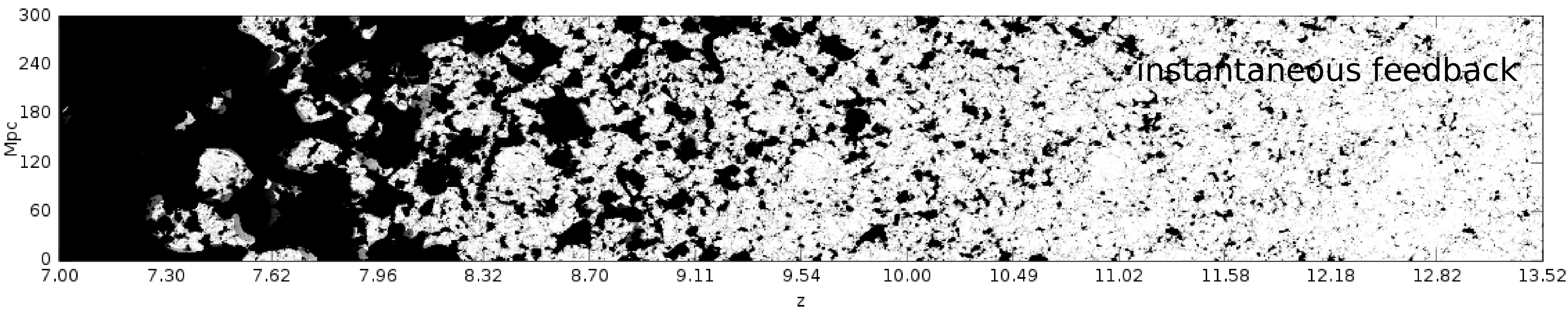}
\includegraphics[width=0.95\textwidth]{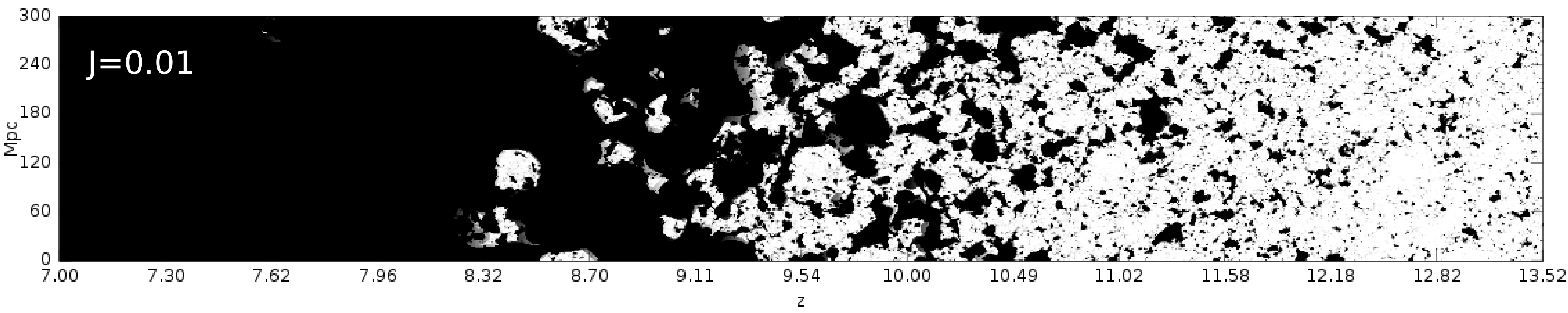}
\includegraphics[width=0.95\textwidth]{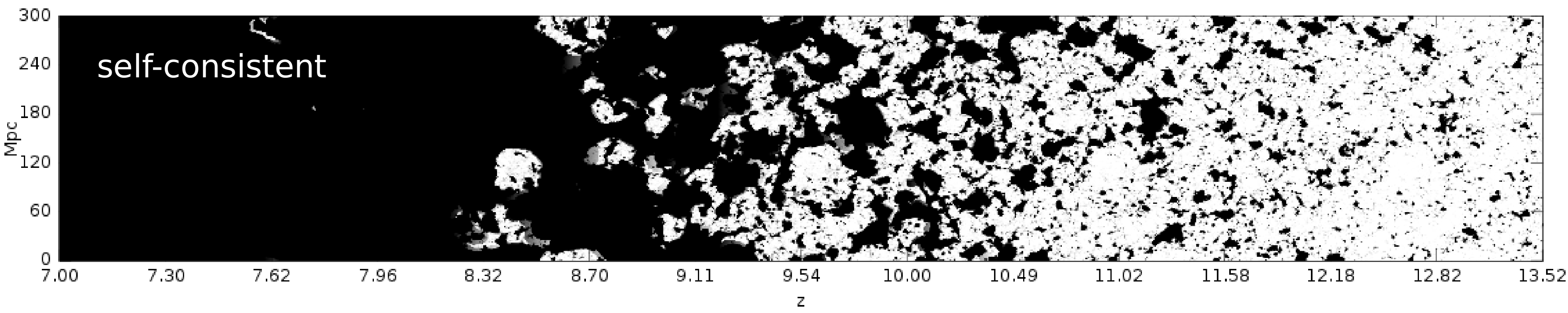}
\includegraphics[width=0.95\textwidth]{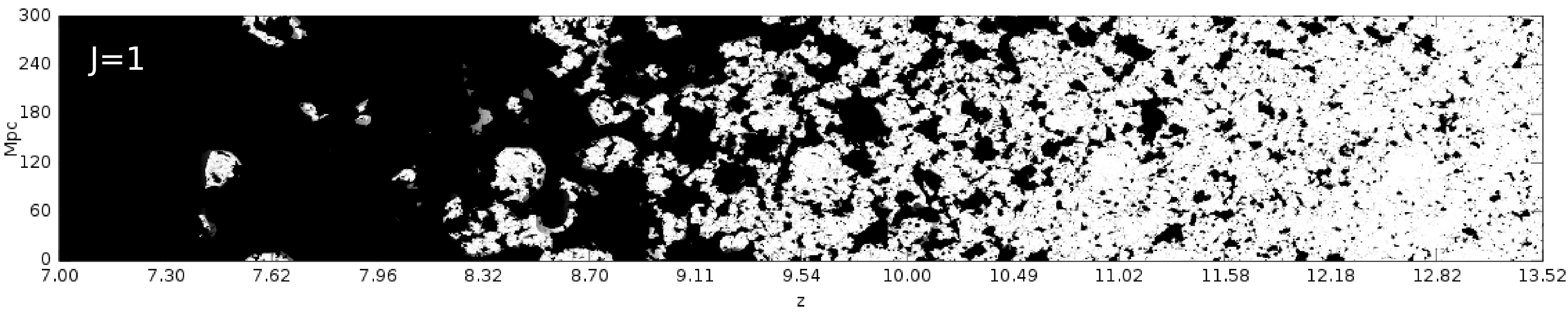}
}
\caption{Slices through our ionization boxes, with HII regions in black and HI regions in white. From top to bottom we show the runs: {\bf NF}, {\bf IF},  {\bf J=0.01}, {\bf SC}, {\bf J=1}.
\label{fig:light_box}
}
\vspace{-1\baselineskip}
\end{figure*}

To model cosmological reionization, we use a parallelized version of the publicly available semi-numerical simulation, \cmfast\footnote{http://homepage.sns.it/mesinger/Sim}.  We generate the IGM density and source fields by: (\textit{i}) creating a 3D Monte Carlo realization of the linear density field in a box with sides $L=300\text{ Mpc}$ and $N=1600^{3}$ grid cells; (\textit{ii}) evolving the density field using the Zel'dovich approximation \citep{Zeldovich70}, and smoothing onto a lower-resolution $N=400^3$ grid; (\textit{iii}) using excursion-set theory \citep{PS74, BCEK91, LC93, ST99} on the evolved density field to compute the fraction of matter collapsed in halos bigger than $M_{\rm min}$, thus contributing to reionization (see \citealt{MF07, MFC11} for a more detailed description of the code).

The ionization field is computed by comparing the integrated number of ionizing photons to the number of baryons, in spherical regions of decreasing radius $R$ (i.e. following the excursion-set approach of \citealt{FHZ04}).  Specifically, a cell located at spatial position and redshift, ($\textbf{x}$, $z$), is flagged as ionized if:
\begin{equation}
\label{eq:ion_crit_coll}
\xi f_{\rm coll}(\textbf{x}, R, z, M_{\rm min})\geq 1
\end{equation}
where $f_{\rm coll}\left(\textbf{x},R,z,M_{\rm min}\right)$ is the fraction of collapsed matter inside a sphere of radius $R$ residing in halos larger than $M_{\rm min}$, and $\xi$ is an ionizing efficiency, defined below. Starting from $R_{\rm max} = R_{\rm mfp} = 30$ Mpc (roughly corresponding to the ionizing photon mean free path in the ionized IGM, $R_{\rm mfp}$, at $z \sim 6$ \citealt{SC10, QOF11}) the smoothing scale, $R$, is decreased, and the criterion in eq. (\ref{eq:ion_crit_coll}) is re-evaluated. At the cell size, the partial ionizations from sub-grid sources are evaluated, including Poisson noise around the mean value of the collapse fraction (e.g. \citealt{MFC11}). This algorithm results in ionization fields which are in good agreement with cosmological radiative transfer algorithms \citep{ZMQT11}.

The ionizing efficiency can be written out as:
\begin{equation}
\label{eq:zeta}
\xi = 30 \bigg(\frac{N_\gamma}{4000}\bigg) \bigg(\frac{f_{\rm esc}}{0.15}\bigg) \bigg(\frac{f_\ast}{0.05}\bigg) \bigg(\frac{1}{1+\bar{n}_{\rm rec}}\bigg) ~ .
\end{equation}
where $f_{\rm esc}$ is the fraction of UV ionizing photons that escape into the IGM, $N_\gamma$ is the number of ionizing photons per stellar baryon, $f_\ast$ is the fraction of galactic gas in stars, and $\bar{n}_{\rm rec}$ is the mean number of recombinations per baryon in the IGM. Our fiducial choice of $\xi=30$ results in an ionization history in agreement with the WMAP observed value of $\tau_e$ \citep{WMAP11}; however we also explore other values below.  Although our models depend only on the product in eq. (\ref{eq:zeta}), we show on the RHS some reasonable values for the component terms.  $N_\gamma \approx 4000$ is expected for PopII stars (e.g. \citealt{BL05_WF}), and studies of the high-redshift Lyman alpha forest suggest $\bar{n}_{\rm rec} \sim 0$ in the diffuse IGM (\citealt{Miralda03, BH07, QOF11}; note that we instead incorporate photon sinks in the form of a mean free path through the ionized IGM, governed by the separation of Lyman limit systems, LLSs).  On the other hand, the parameters $f_{\rm esc}$ and  $f_\ast$ are extremely uncertain in high-redshift galaxies (e.g. \citealt{GKC08, WC09, FL12}), though our fiducial choices are in agreement with high-redshift galaxy luminosity functions (e.g. \citealt{Robertson13}).

{\it The main improvement in this work is that we use the spatially-dependent value of $M_{\rm min}$}, computed according to eq. \ref{eq:M_min} (rather than a homogeneous value $M_{\rm min} = M_{\rm cool}$ as is commonly done).  We keep track of the redshift when each cell was first ionized, $z_{\rm IN}(\textbf{x})$, updating the local value of $M_{\rm crit}(z, z_{\rm IN}, J_{21})$ as reionization progresses (for details on how we compute the inhomogeneous $J_{21}$, see the following section).  Neutral cells are assigned the value $M_{\rm cool}(z)$.  We then use the average value, $\langle M_{\rm min} \rangle_R$ in the ionization criterium of eq. (\ref{eq:ion_crit_coll}).  The drawback of this addition is that it requires ``running-down'' the simulation from high redshifts, rather than being able to independently generate the reionization morphology at a given redshift.

\subsubsection{Incorporating an inhomogeneous, self-consistent UVB}
\label{sec:self_cons}

We also compute the average intensity inside each HII region, for use in eq. (\ref{eq:M_min}).  The mean emissivity (number of ionizing photons emitted into the IGM per second per comoving volume) can be written as:
\begin{equation}
\label{eq:emissivity}
\epsilon \approx f_\ast f_{\rm esc} N_{\gamma} \bar{n}_b \frac{f_{\rm coll}}{t_\ast} \approx \xi \bar{n}_b \frac{f_{\rm coll}}{t_\ast} ~ ,
\end{equation}
where $\bar{n}_b$ is the mean baryon number density inside the HII region, and $t_{\rm *}$ is the star-formation time-scale (which we take to be $t_{*}=0.3\times t_{\rm H}$).  The intensity is proportional to the mean free path inside the HII region (taken to be $\approx R$) multiplied by the emissivity.  If the emissivity is spectrally distributed as $\nu^{-\alpha}$ (eq. \ref{J21}), we can write the average UVB intensity in a given HII region of characteristic size $R$ (in erg s$^{-1}$ Hz$^{-1}$ sr$^{-1}$ proper cm$^{-2}$) as
\begin{equation}
\label{eq:self_cons_J}
\langle J_{\rm 21}\rangle_{\rm HII}=\frac{(1+z)^2}{4\pi} R h\alpha \bar{n}_b \xi \frac{f_{\rm coll}}{t_{*}} ~ .
\end{equation}

Due to the clustering of dark matter halos, the relevant intensity at galaxy locations will be higher than the average UVB intensity in eq. (\ref{eq:self_cons_J}).  Therefore when calculating $M_{\rm min}$, we use $\langle J_{\rm 21}\rangle_{\rm halo, HII}=f_{\rm Jbias}\times\langle J_{\rm 21}\rangle_{\rm HII}$, with $f_{\rm Jbias}=2$, consistent with results from \citet{MD08}.  We also note that the simplification of a uniform UVB inside each HII region seems to be reasonable (see Fig. \ref{fig:J21_evo} and the associated discussion).

Furthermore, this formalism is not fully self-consistent, since our expression for $M_{\rm crit}$ was computed for a time-independent UVB.  Nevertheless, we find that the mean UVB is roughly constant in HII regions, somewhat validating this approximation.
 More importantly, feedback is largely insensitive to the exact value of $J_{21}$ since the gas is always heated to $T\sim 10^{4}\text{ K}$ (specifically, in Paper I we find $M_{\rm crit} \propto J^{0.17}$).

\subsection{Runs}
\label{sec:runs}

The efficient, semi-numerical approach outlined above allows us to explore several scenarios of UVB feedback during reionization.  As already mentioned, our fiducial model assumes $\xi=30$, $T_{\rm cool}=10^{4}\text{ K}$ and $R_{\rm mfp}=30\text{ cMpc}$, though we explore other values throughout.  We have five main prescriptions for implementing UVB feedback:
\begin{itemize}
\item \textbf{No feedback (NF)}: we assume $M_{\rm min}=M_{\rm cool}$, corresponding to a constant virial temperature $T_{\rm cool}$. This model completely neglects UVB feedback, as is commonly done in reionization literature.
\item \textbf{Instantaneous feedback (IF)}: in this extreme model, feedback leads to an {\it instantaneous} transition in $M_{\rm min}$ as soon as the halo is exposed to a UVB, thus ignoring the hydrodynamical response time-scale of the gas. This is the most common prescription\footnote{In simulations, this generally amounts to using the simulation time-step (generally $\approx$10 Myr) as the feedback time-scale.} for UVB feedback (e.g. \citealt{WL03, OM04, IMSP07, WM07, WC07, KC11, AFT12, AISM12}), and we take $M_{\rm min}=M_{\rm cool}$ in HI regions and $M_{\rm min}=10^{9} M_{\odot}$ in HII regions, consistent with previous works. The {\bf IF} and {\bf NF} runs therefore bracket the expected impact of UVB feedback.
\item \textbf{J=1}: we calculate $M_{\rm min}$ with eq. \ref{eq:M_min}, using a constant UVB intensity $J_{\rm 21}=1$ in HII regions. This intensity is larger than estimates from the Lyman alpha forest at $z\sim6$, $J_{\rm 21} \sim 0.1$ \citep{BH07, WB11, CBHB11}.
 Therefore this model places an upper limit to the impact of UVB feedback (which is more physically relevant than the upper limit provided by the {\bf IF} model).
\item \textbf{J=0.01}: like the previous model but with $J_{\rm 21}=0.01$.
\item \textbf{Self-consistent (SC)}: we calculate $M_{\rm min}$ using eq. \ref{eq:M_min}, and we use eq. \ref{eq:self_cons_J} to calculate the inhomogeneous UVB in HII regions.
\end{itemize}

\section{Results}
\label{sec:results}

\subsection{Reionization History}
\label{sec:xHI_evo}

\begin{figure}
\vspace{+0\baselineskip}
{
\includegraphics[width=0.45\textwidth]{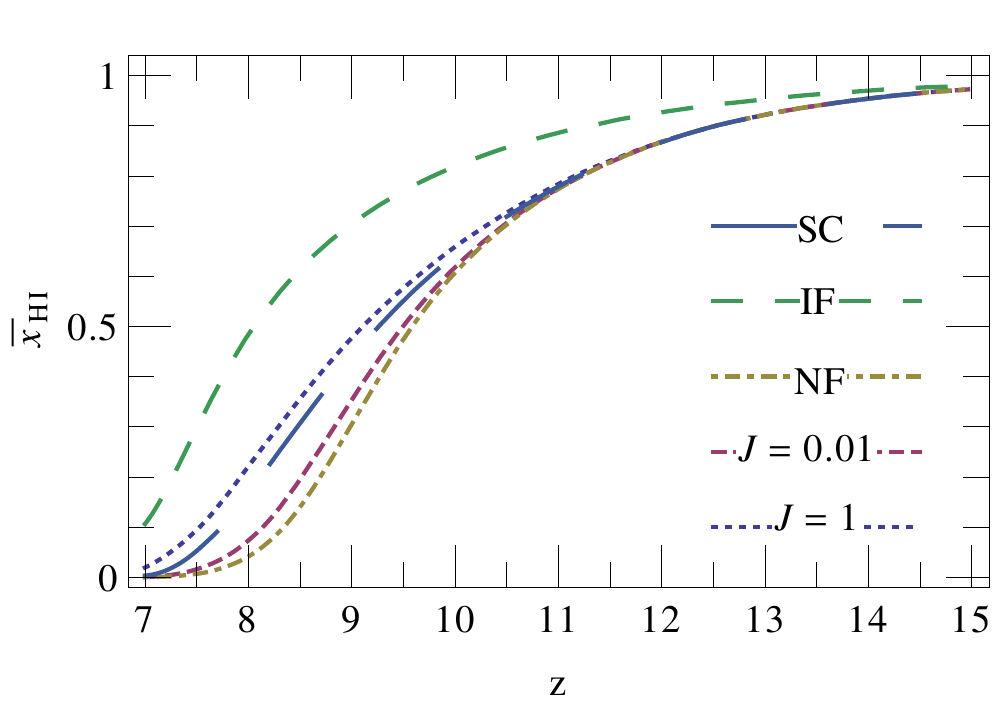}
}
\caption{Evolution of the global neutral fraction $\bar{x}_{\rm HI}$ with different models of radiative feedback. \textit{Dot-Dashed}: no feedback (NF). \textit{Short-dashed}: $J_{\rm 21}=0.01$ inside HII regions. \textit{Dotted}: $J_{\rm 21}=1$ inside HII regions. \textit{Dashed}: instantaneous feedback (IF). \textit{Long-dashed}: self-consistent model (SC).
\label{fig:xHI_evo}
}
\vspace{-1\baselineskip}
\end{figure}

In Fig. \ref{fig:light_box}, we show slices through the ionization boxes of our five fiducial models:
{\bf NF}, {\bf IF},  {\bf J=0.01}, {\bf SC}, {\bf J=1} ({\it top to bottom}). The corresponding electron-scattering optical depths are: $\tau_{\rm e}=0.088$, $\tau_{\rm e}=0.073$, $\tau_{\rm e}=0.087$, $\tau_{\rm e}=0.085$, $\tau_{\rm e}=0.084$\footnote{UV-driven reionization is ``inside-out'' on large-scales, with overdensities hosting the biased galaxies being the first to ionize.  \citet{MFS12} noted that this correlation between the density and ionization fields results in a $\approx4$\% higher $\tau_{\rm e}$ than what would be expected assuming the two fields are uncorrelated: $\langle x_i \times n_b \rangle \neq \langle x_i \rangle \times \langle n_b \rangle$.
For example our self-consistent model has $\tau_{\rm e}=0.085$ instead of $\tau_{\rm e}=0.083$ as would be expected ignoring this correlation.} consistent at $1\sigma$ with WMAP results \citep{WMAP11}.

As expected, UVB feedback delays reionization and results in a more uniform distribution of HII regions (e.g. \citealt{QLZD07, IMSP07}).  However, our self-consistent model is not remarkably different from the one which ignored UVB feedback altogether ({\bf NF}).  Most striking is the effect of an unphysical instantaneous transition in $M_{\rm min}$: the {\bf IF} model dramatically over-predicts the importance of UVB feedback even with respect to the extreme {\bf J=1} model.  We can therefore immediately surmise that {\it the delay in the hydrodynamic response of the baryons to the UVB is fundamental in properly modeling UVB feedback}.

This qualitative result is confirmed by the evolution of the global neutral fraction $\bar{x}_{\rm HI}$, shown in Fig. \ref{fig:xHI_evo}. As expected, the self-consistent model shows intermediate features between the {\bf J=0.01} and {\bf J=1} models. Since the transition of $M_{\rm min}$ is not instantaneous, only the late stages of reionization are delayed.  Comparing the {\bf SC} and {\bf NF} curves, we see that UVB feedback delays the later stages of reionization by $\Delta z \lsim 0.5$.  On the other hand, this delay is significantly overestimated by a model with instantaneous feedback: $\Delta z \approx 1.5$.

\subsection{Reionization Morphology}

\begin{figure*}
\vspace{+0\baselineskip}
{
\includegraphics[width=0.33\textwidth]{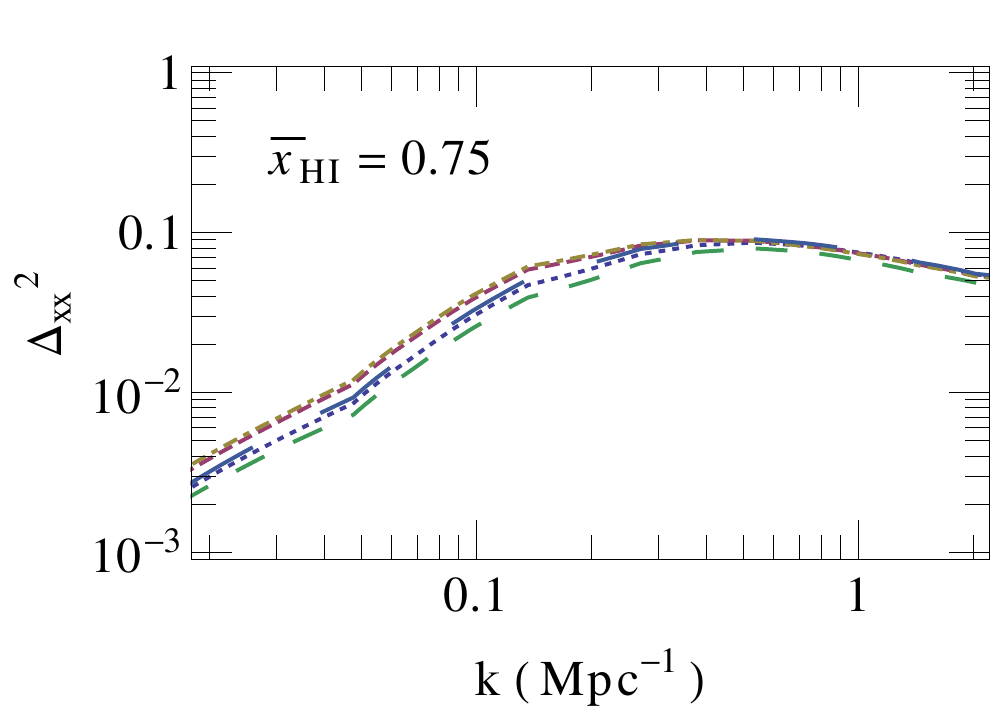} 
\includegraphics[width=0.33\textwidth]{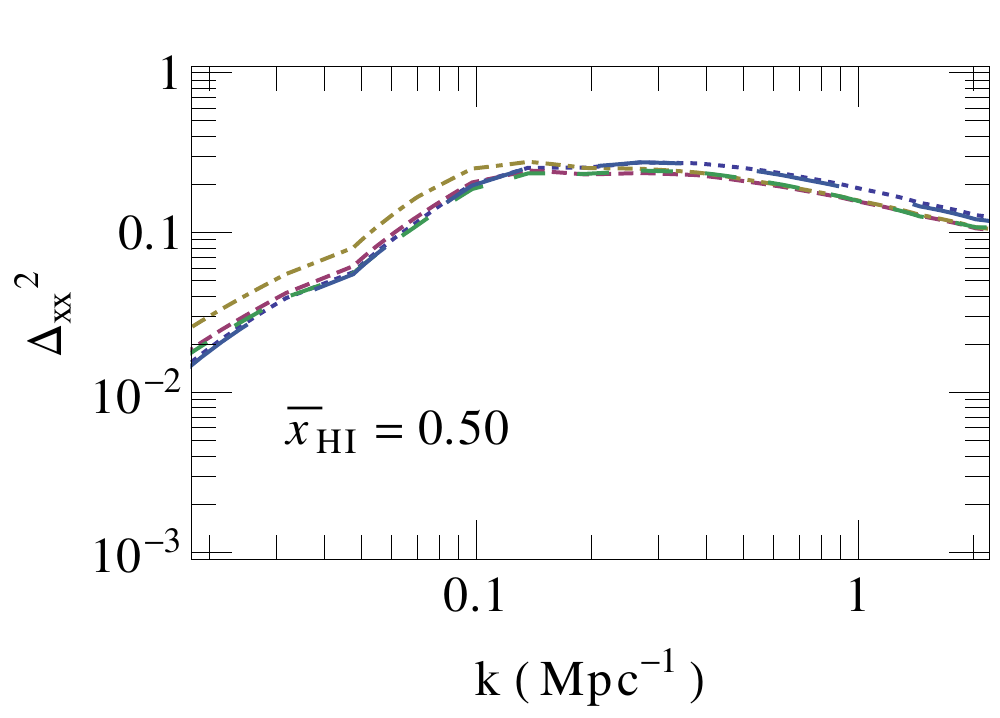} 
\includegraphics[width=0.33\textwidth]{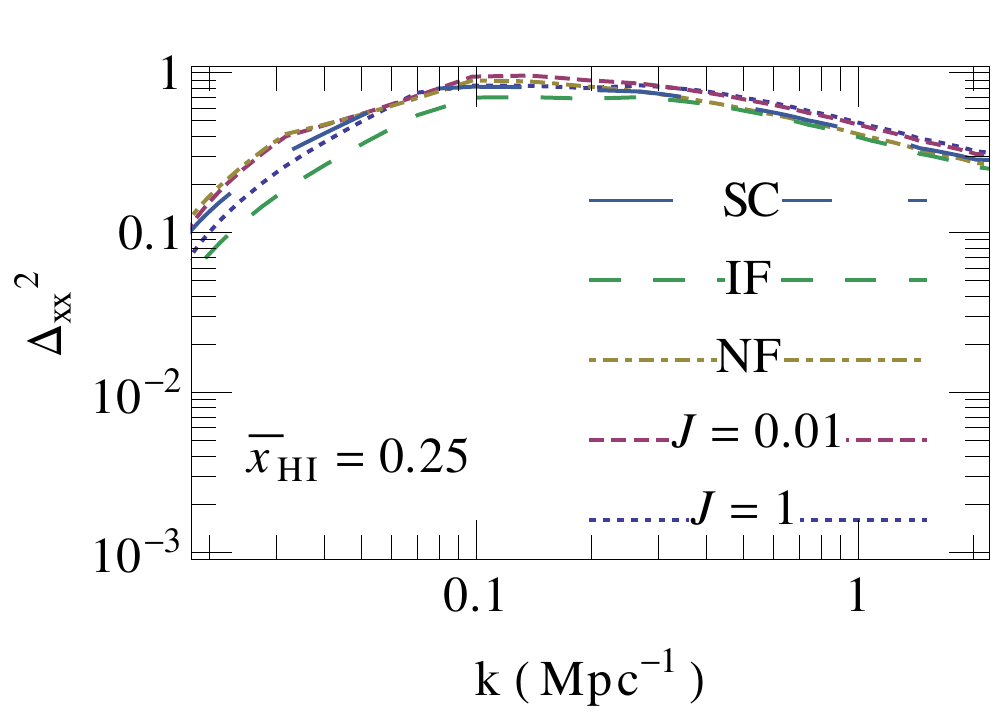}
}
\caption{Power spectrum $\Delta_{\rm xx}^{2}$ of the ionization field at $\bar{x}_{\rm HI}=0.75$ (\textit{left}), $\bar{x}_{\rm HI}=0.50$ (\textit{middle}) and $\bar{x}_{\rm HI}=0.25$ (\textit{right}) with different models of UVB feedback. \textit{Dot-Dashed}: {\bf NF}. \textit{Short-dashed}: {\bf J=0.01}. \textit{Dotted}: {\bf J=1}. \textit{Dashed}: {\bf IF}. \textit{Long-dashed}: {\bf SC}.
\label{fig:pow_spec}
}
\vspace{-1\baselineskip}
\end{figure*}

\begin{figure*}
\vspace{+0\baselineskip}
{
\includegraphics[width=0.33\textwidth]{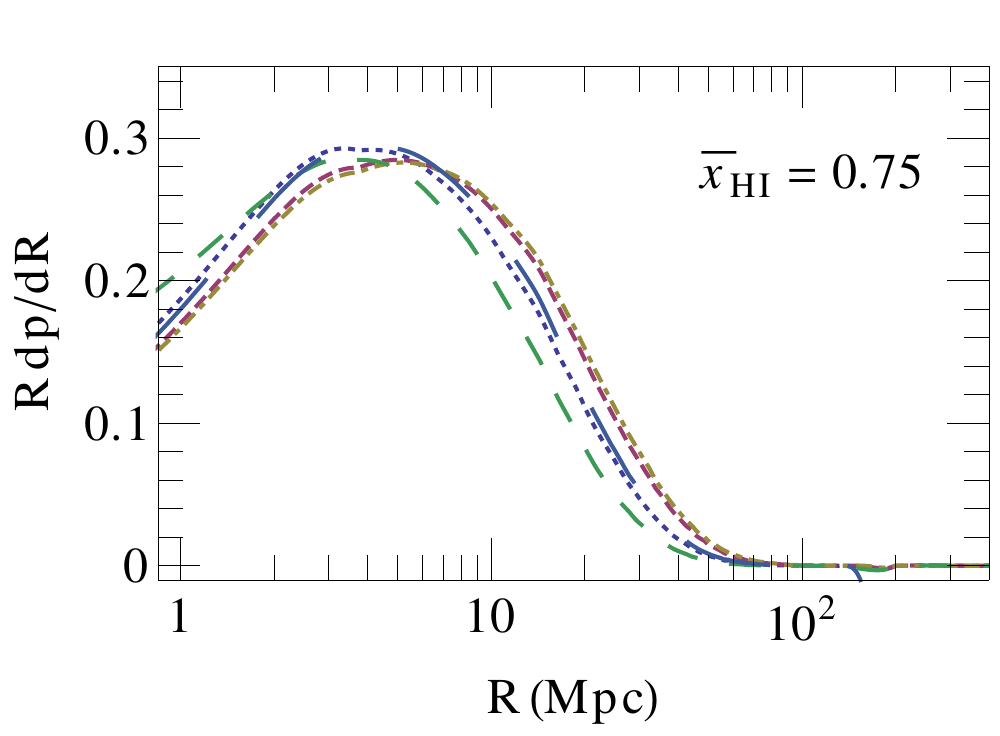}
\includegraphics[width=0.33\textwidth]{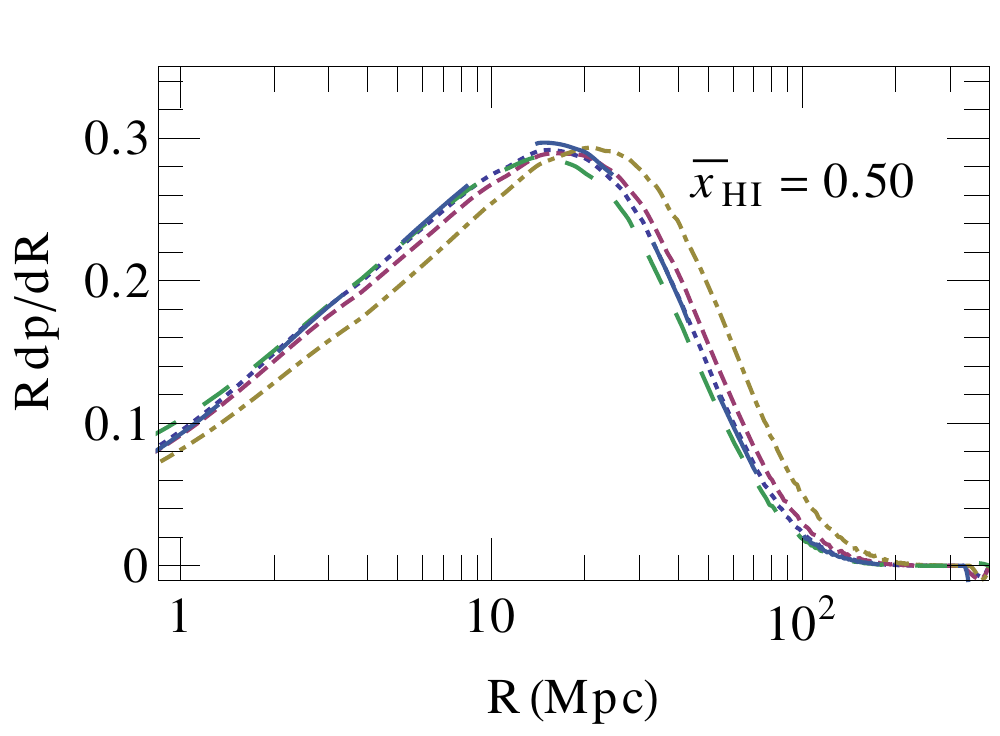}
\includegraphics[width=0.33\textwidth]{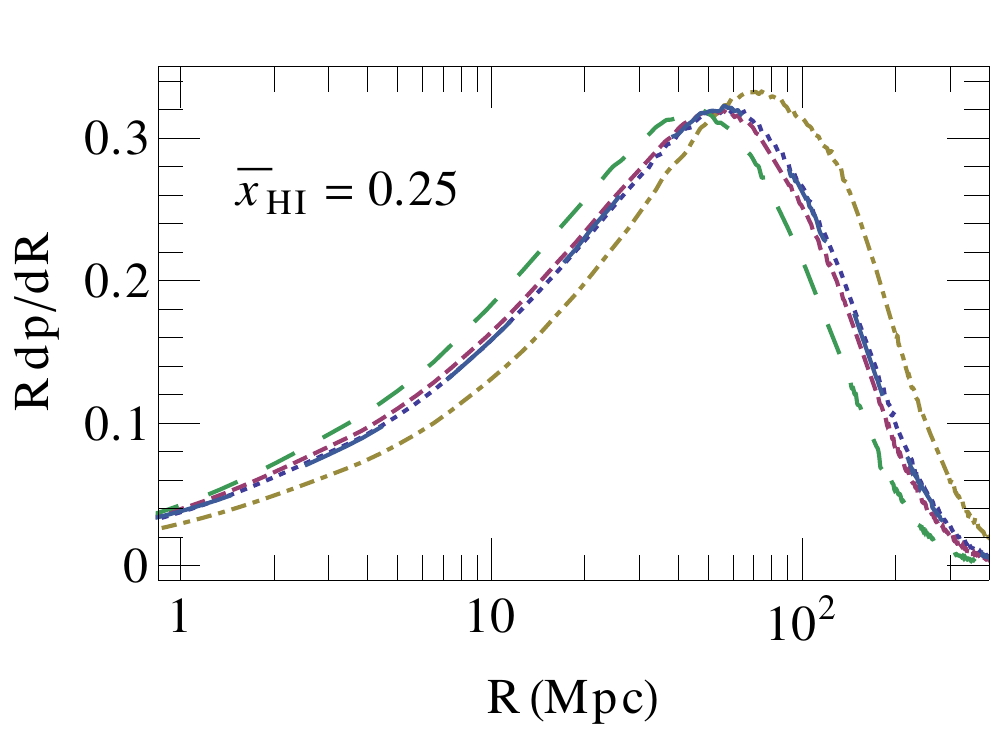}
}
\caption{Size distribution of the HII regions at $\bar{x}_{\rm HI}=0.75$ (\textit{left}), $\bar{x}_{\rm HI}=0.50$ (\textit{middle}) and $\bar{x}_{\rm HI}=0.25$ (\textit{right}) with different models of UVB feedback. \textit{Dot-Dashed}: {\bf NF}. \textit{Short-dashed}: {\bf J=0.01}. \textit{Dotted}: {\bf J=1}. \textit{Dashed}: {\bf IF}. \textit{Long-dashed}: {\bf SC}.
\label{fig:size_distr}
}
\vspace{-1\baselineskip}
\end{figure*}

We now proceed to quantify the impact of UVB feedback on the morphology of reionization. In Fig. \ref{fig:pow_spec} we show the ionization power spectrum ($\Delta^2_{\rm xx} \equiv k^3/(2\pi^2 V) ~ \langle|\delta_{\rm xx}|^2\rangle_k$, with $\delta_{\rm xx}=x_{\rm HI}/\avenf - 1$),
 while in Fig. \ref{fig:size_distr} we show the size distributions of the HII regions\footnote{We calculate the size distributions following the prescription in \citet{MF07}. Namely, we randomly choose an ionized cell and then 
record the comoving distance to the edge of the HII region in a random direction. Then we define the size distribution as $Rdp/dR$, where $p$ is the probability that this distance is between $R$ and $R+dR$. Technically we define the edge of an HII region as the first cell with $x_{\rm HI}\geq 0.5$. This choice of threshold can affect the size distributions somewhat; however, we are interested in the relative difference between the models, which is much more robust to this choice \citep{FMAS11}.}.  Panels correspond to different stages of reionization:
 $\bar{x}_{\rm HI}=0.75, 0.50, 0.25$, ({\it left to right}). We compare our fiducial models of UVB feedback: no feedback (yellow dot-dashed line); $J_{\rm 21}=0.01$ (magenta short-dashed line); $J_{\rm 21}=1$ (blue dotted line); instantaneous feedback (green dashed line); self-consistent (blue long-dashed line).

As expected from Fig. \ref{fig:light_box}, we see that ignoring feedback over-predicts the large-scale ionization power by tens of percent.   The same trend can be seen in the HII size distributions, with the peak in the {\bf NF} model being shifted by $\approx50$\% with respect to the {\bf SC} model.  The {\bf IF} model on the other hand over-estimates these feedback effects by a comparable amount. We therefore confirm earlier predictions that UVB feedback results in smaller, more uniform ionized regions \citep{QLZD07}.  This is easy to understand: the largest HII regions correspond to the most biased locations of the density field, and are therefore ionized earliest.  It is in these regions that UVB feedback has had the most time to quench star-formation.  Smaller, late-forming HII regions are less affected, and therefore more abundant at fixed $\avenf$.

Nevertheless, the impact of UVB feedback is relatively minor, when compared at fixed $\avenf$: the power spectra of all models agree to within a factor of $\lsim 2$.  This is due to the fact that the halo bias evolves only weakly over the mass scales relevant to UVB feedback; hence morphology at a given $\avenf$ is relatively robust \citep{QLZD07}.  In other words, reionization morphology is much more sensitive to $\avenf$ than it is to UVB feedback effects.  This implies that previous predictions for reionization morphology (at fixed $\avenf$) are not particularly sensitive to this source of astrophysical uncertainty.

\subsection{Evolution of $\textbf{M}_{\textbf{min}}$}
\label{sec:M_min_evo}

\begin{figure}
\vspace{+0\baselineskip}
{
\includegraphics[width=0.45\textwidth]{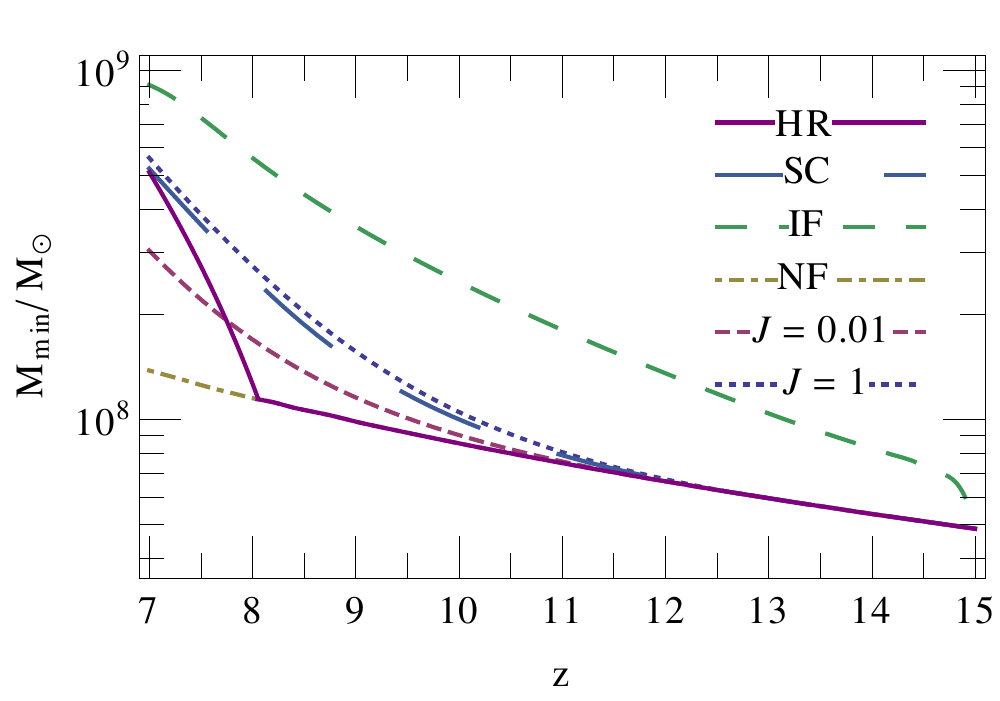}
}
\caption{Average value of $M_{\rm min}$ versus redshift. \textit{Dot-dashed}: no feedback ({\bf NF}). \textit{Short-dashed}: our model with $J_{\rm 21}=0.01$. \textit{Dotted}: our model with $J_{\rm 21}=1$. \textit{Dashed}: instantaneous feedback ({\bf IF}). \textit{Long-dashed} self-consistent model ({\bf SC}). \textit{Solid}: for comparison we show $M_{\rm min}$ (eq. \ref{eq:M_min}) assuming a homogeneous reionization ({\bf HR}) at $z_{\rm IN}=9.3$ (corresponding to the mid-point of reionization in the {\bf SC} model).
\label{fig:av_mc}
}
\vspace{-1\baselineskip}
\end{figure}

The results of the previous sections can be more readily understood by looking at the redshift evolution of the average value of the minimum halo mass hosting star-forming galaxies, $\bar{M}_{\rm min}$, shown in Fig. \ref{fig:av_mc}, using the same line styles as above.  This quantity is also fundamental in semi-analytical models of galaxy formation.  

By construction, the $\bar{M}_{\rm min}$ curve in the {\bf NF} model simply follows the $T_{\rm vir} = 10^4$ K isotherm.  The curve corresponding to the {\bf SC} model is closer to the {\bf J=1} model than the {\bf J=0.01} one (see Fig. \ref{fig:J21_evo} and associated discussion). It begins ``pealing off'' the $T_{\rm vir} = 10^4$ K isotherm at $z\sim10$, when reionization is already well underway ($\avenf\sim0.6$).  In contrast, $\bar{M}_{\rm min}$ in the {\bf IF} model immediately starts increasing as reionization progresses, since in that model the average minimum halo mass is simply $\bar{M}_{\rm min} = \avenf M_{\rm cool} + (1-\avenf)10^9 M_\odot$.  Because it ignores the delay associated with UVB feedback, the {\bf IF} model over-predicts $\bar{M}_{\rm min}$ by a factor of $\sim$2--3, with respect to the {\bf SC} model.

For comparison we also show the evolution of $\bar{M}_{\rm min}$ assuming a homogeneous reionization at $z_{\rm IN}=9.3=z_{\rm re}$, corresponding to the redshift when $\bar{x}_{\rm HI}=0.5$ in the self-consistent model (magenta solid line). This formula under-predicts $\bar{M}_{\rm min}$ by a factor of $\sim 2$ around the mid-point of reionization, highlighting that the {\it scatter} in the reionization redshifts of galaxies can be important in regulating star formation (e.g. \citealt{ABAW09}).  This homogeneous reionization model further highlights the delay in the hydrodynamic response of the gas to the UVB: $M_{\rm crit}$ only surpasses $M_{\rm cool}$ at $z \approx 8$, even though the ionizing background is effectively turned on at $z_{\rm re}=9.3$ (see also \citealt{OGT08}).

It is natural to approximate this delay with the sound-crossing time-scale, $t_{\rm sc} = 2R_{\rm vir}/c_{\rm s}(10^4\text{ K})$ (which is comparable to the photo-evaporation time-scale, e.g. \citealt{HAM01, SIR04}).  For a halo of mass $M$, the sound-crossing time-scale can be approximated as (c.f. \citealt{SIR04}):
\begin{equation}
\label{eq:t_sc}
t_{\rm sc} \approx 200 \text{Myr} \left( \frac{M}{10^8 M_\odot} \right)^{1/3} \left( \frac{1+z}{10} \right)^{-1} \left( \frac{\Omega_{\rm m} h^2}{0.15}\right)^{-1/3} ~.
\end{equation}

It would therefore be reasonable to expect $\bar{M}_{\rm min}$ to evolve from $M_{\rm cool}$ at high redshifts to some $M_0(T_{\rm vir}\approx$ const) at low redshifts, with the transition occurring at $z \sim z_{\rm re} - \Delta z_{\rm sc}$ [where $\Delta z_{\rm sc}$ is the redshift interval corresponding to eq. (\ref{eq:t_sc})], roughly over a time-scale scaling with the duration of reionization, $\Delta z_{\rm re}$.  Indeed, we find that our results for $\bar{M}_{\rm min}$ are well fitted by:
\begin{equation}
\label{eq:MC}
\bar{M}_{\rm min}(z)=M_{\rm cool}\times\left(\frac{M_{\rm 0}}{M_{\rm cool}}\right)^g ~ ,
\end{equation}
\noindent with the transition function
\begin{equation}
g\left(z\right)=\frac{1}{1+\exp\left[\frac{z-(z_{\rm re}-\Delta z_{\rm sc})}{\Delta z_{\rm re}}\right]} ~ .
\end{equation}
We take $M_{\rm 0}(z)$ to correspond to a fixed virial temperature $T_{\rm 0}=5\times 10^{4}\text{ K},$\footnote{We caution that our simple 1D simulations from Paper I are not well suited for low-redshift structure formation, where mergers and other 3D effects can be important.  Hence, here we use a lower value of $M_{\rm 0}$ than implied by extrapolating eq. (\ref{eq:critical_mass_3}) to $z=0$; our choice is roughly consistent with cosmological simulations of homogeneous reionization \citep{OGT08}.} and define $\Delta z_{\rm re}$ to correspond to the redshift interval between $\bar{x}_{\rm HI}=0.6$ and $\bar{x}_{\rm HI}=0.4$.

\begin{figure}
\vspace{+0\baselineskip}
{
\includegraphics[width=0.45\textwidth]{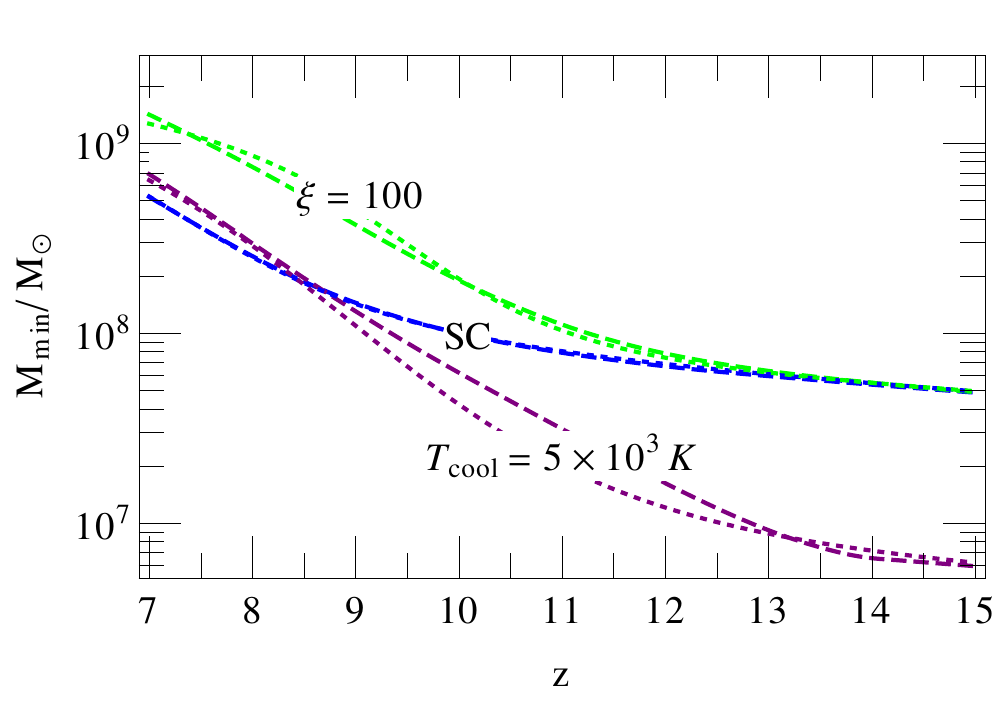}
}
\caption{Average value of $M_{\rm min}$ versus redshift. We compare the self-consistent model with the standard parameter choice (SC), with $T_{\rm cool}=5\times 10^{3}\text{ K}$ and with $\xi=100$. \textit{Dashed}: results of the simulation. \textit{Dotted}: analytic expression in eq. \ref{eq:MC}.
\label{fig:av_mc_2}
}
\vspace{-1\baselineskip}
\end{figure}

In Fig. \ref{fig:av_mc_2} we compare this analytic approximation (dotted lines) with the results from our simulation (dashed lines). We show the self-consistent model with the standard choice of the parameters (in this case the analytic formula and the result of simulation are indistinguishable).  To test the robustness of the above approximation, we also include two "extreme" models: (i) with molecular hydrogen cooled halos significantly contributing to reionization: $T_{\rm cool}=5\times 10^{3}\text{ K}$ ({\it magenta curves}); (ii) with a significantly higher emissivity $\xi=100$ ({\it green curves}). These runs correspond to $z_{\rm re}$ = (9.3, 9.9, 12.0), $\Delta z_{\rm re}$ = (1.0, 1.3, 0.8) and $\Delta z_{\rm sc}$ = (2.0, 0.9, 2.5), for the fiducial, $T_{\rm cool}=5\times 10^{3}\text{ K}$, and $\xi=100$ models, respectively. 
Even in the extreme cases, our formula from eq. (\ref{eq:MC}) provides a good fit.  {\it This general, physically-motivated expression facilitates implementing UVB feedback from an inhomogeneous reionization into models of galaxy formation.}

\subsection{UVB Evolution}

\begin{figure}
\vspace{+0\baselineskip}
{
\includegraphics[width=0.45\textwidth, height=5.5cm]{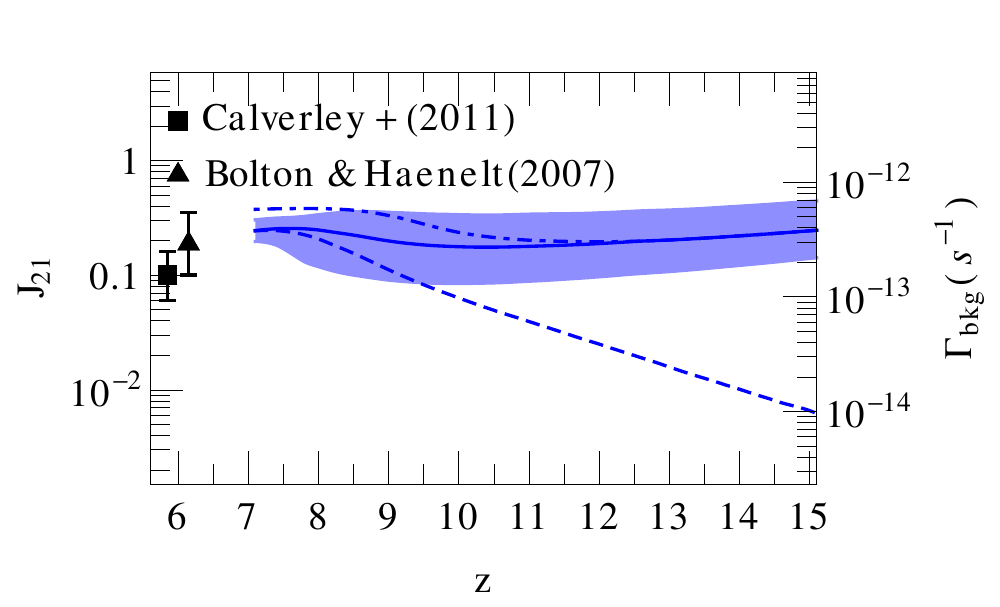}
}
\caption{Evolution of the UVB intensity $J_{\rm 21}$ (left-hand scale) and of the corresponding photoionization rate per baryon $\Gamma_{\rm bkg}$ (right-hand scale). \textit{Solid}: average intensity $\langle J\rangle_{\rm HII}$ within HII regions in our {\bf SC} model; the shaded region corresponds to the spread among HII regions. \textit{Dashed}: average intensity $\langle J\rangle_{\rm V}$ in the entire simulation box. \textit{Dot-dashed}: evolution of $\langle J\rangle_{\rm HII}$ when feedback is neglected ({\bf NF}). For comparison we show $\Gamma_{\rm bkg}$ at $z\approx6$ (offset for clarity) as estimated from the Ly$\alpha$ forest \citep{BH07, CBHB11}.
\label{fig:J21_evo}
}
\vspace{-1\baselineskip}
\end{figure}

In addition to the evolution of HII regions, UVB feedback affects the UVB itself.  As star-formation is suppressed in vulnerable halos, the UVB decreases (with respect to the case without feedback). Our formalism explicitly computes the inhomogeneous UVB intensity, as described in \S \ref{sec:self_cons}.  It therefore provides a useful prediction of the evolution of $J_{21}$ during reionization.

The UVB can be determined from measurements of the Lyman alpha forest in high-redshift quasar spectra.  The ionization rate per baryon is found to be remarkably constant from $z\approx 2 \rightarrow 5$, $\Gamma_{\rm bkg}\approx10^{-12}$ s$^{-1}$ \citep{BH07, F-G08}.  Then from $z\approx 5 \rightarrow 6$, there is evidence of a drop by a factor of few, though the uncertainty in the measurement is large at these high-redshifts when the Ly$\alpha$ forest begins to saturate \citep{BH07, WB11, CBHB11}.  Such a drop could be caused by either an incomplete reionization \citep{Mesinger10, MMF11}, and/or by a rapid evolution in the abundance of LLSs \citep{FO05, QOF11}.\footnote{Note that a large drop in the ionizing background must involve the photo-evaporation of LLSs \citep{QOF11}, since these systems regulate the progress of reionization during its final stages (e.g. \citealt{FM09, CMMF11, AA12}).  We caution that our models do not incorporate an evolving mean free path in ionized regions. Therefore, they likely underestimate the rise in the UVB in the late stages and following reionization.  However, the relative impact of feedback is more robust to this uncertainty.}

In Fig. \ref{fig:J21_evo} we show the redshift evolution of the UVB in our {\bf SC} (solid curve) and {\bf NF} (dot-dashed curve) models ($J_{\rm 21}$/$\Gamma_{\rm bkg}$ on the left/right vertical scales), together with $z\approx6$ observational estimates by \citet{BH07} and \citet{CBHB11}.  It is important to note that during UV-driven reionization, {\it the UVB is expected to be bi-modal}, with $\langle J_{21} \rangle_{\rm HII}$ inside HII regions and $J_{21}\sim0$ inside neutral regions, such that the volume-averaged UVB intensity is $\langle J_{21} \rangle_{\rm V} \sim (1-\avenf)\langle J_{21} \rangle_{\rm HII}$.  This distinction is generally not made in Lyman forest studies, which thereby implicitly assume that reionization is over (though see \citealt{GCF06} and \citealt{MMF11}), or that it is driven by X-rays with long mean free paths (e.g. \citealt{Oh01}).  To highlight this point, in Fig. \ref{fig:J21_evo} we also show the evolution of $\langle J_{21} \rangle_{\rm V}$ in the {\bf SC} model (dashed curve).

From Fig. \ref{fig:J21_evo}, we see that the average UVB intensity within HII regions in the {\bf SC} model is quite flat, with a value of $\langle J\rangle_{\rm HII}\simeq 0.2$, consistent with observational estimates at $z\sim6$.\footnote{Choosing a lower/higher value of the ionizing efficiency, $\xi$, would delay/advance reionization and decrease/increase the resulting photoionization rate.  The observational data still allow for a relatively broad range of $\xi$.}  Comparing the solid and dot-dashed curves, we see that feedback in our {\bf SC} model suppresses the UVB by a factor of $\lsim 2$ towards the end of reionization.  This factor is modest, compared to the uncertainties in the model parameters and observational measurements.

The shaded region in Fig. \ref{fig:J21_evo} corresponds to the 1$\sigma$ scatter in $\langle J_{21} \rangle_{\rm HII}$.  This spread is roughly a factor of $\sim2$ during most of reionization, decreasing to $\lsim50$\% in the final overlap stages.  Such a spread in the HII-region averaged intensity is roughly comparable to the overall spread in the intensity, as estimated by \citet{MD08} and  \citet{MF09} who summed contributions from individual galaxies.  This supports our formalism in \S \ref{sec:self_cons}, and the corresponding simplification of using a roughly uniform ionizing background inside each HII region.

\subsection{Was Reionization ``Self-Regulated''?}

Using the assumption of an instantaneous change in $M_{\rm min}$ inside ionized regions (i.e. our {\bf IF} model), many works claim that UVB feedback has a dramatic impact on reionization history and the corresponding star-formation rate (e.g. \citealt{WL03, OM04, IMSP07, WM07, WC07, KC11, AFT12, AISM12}).  Some works dub this effect to be ``self-regulation'', since when HII regions form, their subsequent growth is impaired.
However, we show above that under more realistic prescriptions of UVB feedback, this effect is relatively minor (see also \citealt{MD08}).  This qualitative claim is easy to understand: the time-scales associated with UVB feedback are comparable to the time-scale of reionization.  Thus in the regimes we study, there is not enough time for UVB feedback to efficiently delay reionization.  We now ask the question, {\it ``what are the physically-motivated regimes in which UVB feedback can in fact self-regulate reionization?''}  Naively one might expect that the strength of UVB feedback can be increased by (\textit{i}) a higher UVB intensity; (\textit{ii}) a more extended reionization; (\textit{iii}) a larger contribution of small halos, more susceptible to UVB feedback.  We explore each of these in turn.

\begin{figure}
\vspace{+0\baselineskip}
{
\includegraphics[width=0.45\textwidth]{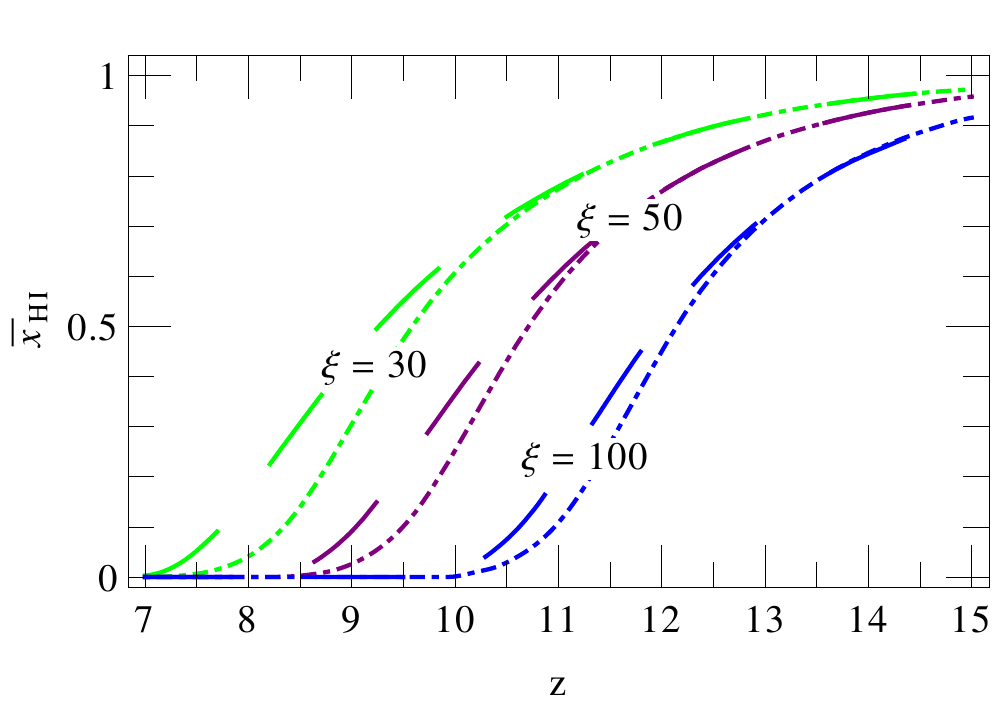}
}
\caption{Evolution of the global neutral fraction $\bar{x}_{\rm HI}$ with different ionizing efficiencies $\xi=30, 50, 100$. \textit{Dot-Dashed}: no feedback. \textit{Long-dashed}: self-consistent model.
\label{fig:self_reg}
}
\vspace{-1\baselineskip}
\end{figure}

In Fig. \ref{fig:self_reg} we show the reionization history in models with different ionizing photon efficiencies: $\xi=$30, 50, 100 ({\it left to right pairs of curves}).
We compare the evolution when UVB feedback is neglected ({\bf NF}; dot-dashed line) with our self-consistent approach ({\bf SC}; long-dashed line).  In fact, increasing the UVB intensity ($J_{\rm 21}\propto\xi$), results in smaller feedback effects.  This is because the dependence of $M_{\rm min}$ on $J_{\rm 21}$ is weak in our model ($M_{\rm crit} \propto J_{\rm 21}^{0.17}$), while for high values of $\xi$, reionization proceeds faster. Therefore as the ionizing emissivity is increased, the decrease in the duration of reionization is more relevant than the delay resulting from the increase in $M_{\rm crit}$, and the evolution of reionization is driven by the emissivity.

\begin{figure}
\vspace{+0\baselineskip}
{
\includegraphics[width=0.45\textwidth]{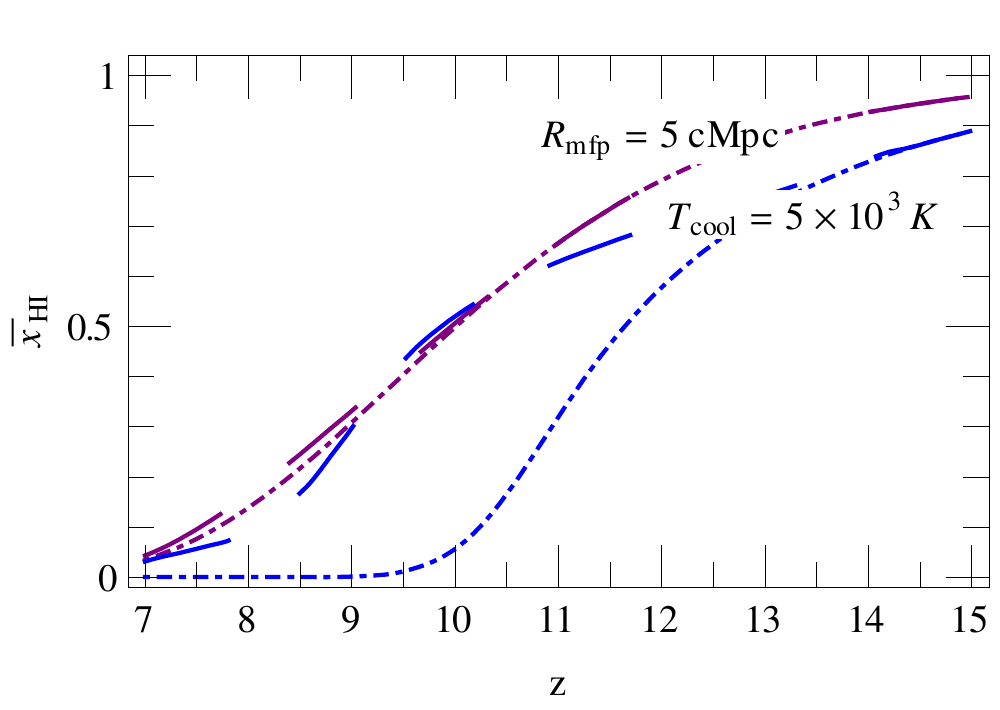}
}
\caption{Evolution of the global neutral fraction $\bar{x}_{\rm HI}$ with $R_{\rm mfp}=5\text{ cMpc}$ and with $T_{\rm vir}=5\times 10^{3}\text{ K}$. \textit{Dot-Dashed}: no feedback. \textit{Long-dashed}: self-consistent model.
\label{fig:extreme_models}
}
\vspace{-1\baselineskip}
\end{figure}

We next attempt to extend the duration of reionization, in order to give the UVB time to significantly suppress the gas content of galaxies. The most efficient way of significantly extending reionization\footnote{Note that we already include star formation in halos down to the atomic-cooling threshold.  The fractional abundance of these halos evolves more slowly than that of more massive halos, expected to have more efficient star-formation (e.g. \citealt{Lidz07}).  Therefore increasing the duration of reionization by changing the population of host halos is only possible if we include molecularly-cooled galaxies (see below).} is to decrease the ionizing photon mean free path through the ionized IGM, $R_{\rm mfp}$ (e.g. \citealt{AA12, MQS12}). Our fiducial model assumes $R_{\rm mfp}=30$ Mpc, consistent with observational and theoretical estimates at $z\approx6$ (e.g. \citealt{SC10, QOF11}). In Fig. \ref{fig:extreme_models} we show the reionization history of our fiducial model, but decreasing $R_{\rm mfp}$ to 5 Mpc.  Although the reionization history is significantly extended, UVB feedback has a negligible effect, resulting in a delay of $\Delta z \lsim 0.1$.  This is because the effective horizon imposed by $R_{\rm mfp}$ restricts feedback to a very limited volume surrounding each galaxy.  Reionization proceeds in a disjoint manner: with tiny, isolated HII regions emerging around newly-forming galaxies \citep{AA12, MQS12}.  Such a reionization scenario is not affected by UVB feedback.

Finally, we turn to our point (iii): a larger contribution of small halos, more susceptible to UVB feedback.  So far, we have considered halos down to $T_{\rm cool}=10^{4}\text{ K}$: this is conservatively small, since winds and internal feedback can suppress star-formation inside such halos (e.g. \citealt{SH03, PS09, FDO11}).  However, some models argue for a significant contribution of galaxies hosted by even smaller halos (so-called minihalos), in which gas collapses through molecular cooling channels.  In Fig. \ref{fig:extreme_models}, we show the reionization history of such a toy model which allows halos down to $T_{\rm cool}=5\times 10^{3}\text{ K}$ to emit ionizing photons with the same fiducial efficiency $\xi=30$.  In this extreme case, UVB feedback has a dramatic impact on reionization history: delaying the advanced stages of reionization by $\Delta z\sim$ 1.5--2, and resulting in a slow evolution of $\avenf$ during the first half of reionization.  This result is qualitatively consistent with the recent results by \citet{Ahn12} (who assumed instantaneous UVB feedback, as well as instantaneous feedback from an H$_2$-disassociative background; the later results in an even flatter, self-regulated regime during the early stages).\footnote{We caution that such simple models very likely overestimate the ionizing efficiencies of minihalos, by ignoring the chemical and mechanical feedback from the first stars (e.g. \citealt{TFS07, Whalen08SN}), X-ray pre-heating of the IGM (e.g. \citealt{RO04}), as well as the baryon-DM velocity offset (e.g. \citealt{TH10}).  Simulations predict that these mechanisms are very efficient in suppressing star-formation in minihalos.}

We therefore conclude that {\it reionization is self-regulating only if star-formation is efficient inside molecularly-cooled galaxies}. We caution however that our implementation of UVB feedback from Paper I was calibrated only for atomic cooled halos.  Furthermore, our models assume that the star-formation rate is proportional to the total amount of gas inside halos.  We do not model the ability of gas to cool and form stars, which can be dramatically suppressed in minihalos by an H$_2$-disassociative background {\it before the bulk of reionization} (assuming modest values for their ionizing emissivities; e.g. \citealt{HRL97}).

\section{Conclusions}
\label{sec:concl}

Cosmic reionization inhomogeneously heated the IGM, affecting its cooling properties and photo-evaporating gas from the outskirts of galaxies.  As a result, small dwarf galaxies inside the ionized patches would have a reduced gas reservoir for star-formation.  This process involves a huge dynamic range; therefore large-scale reionization simulations generally either ignore this effect or include it with simple, fairly ad-hoc prescriptions.

In this work, we use a tiered approach to study the impact of UVB feedback on reionization.  Combining parameterized results for the minimum halo mass capable of hosting star-forming galaxies ($M_{\rm min}$, see Paper I) with semi-numerical simulations of reionization, we present large-scale (300 Mpc) reionization simulations which include UVB feedback in a physically-motivated manner.  In our models, the ionizing emissivity of galaxies is assumed to be proportional to their baryonic content, and depends on the local values of the reionization redshift and the UVB intensity.  These additions will be included in an upcoming version update of \cmfast.

UVB feedback delays the end stages of reionization by $\Delta z \lsim 0.5$, suppresses the large-scale ionization power by tens of percent (at fixed $\avenf$), and decreases the mean photoionization rate by a factor of $\lsim2$ towards the end of reionization.  These effects are quite modest, when compared to current astrophysical and observational uncertainties.  By contrast, the popular assumption of an instantaneous change in $M_{\rm min}$ results in a much larger effect, with a $\Delta z \approx 1.5$ delay in reionization, and a factor of few higher average $M_{\rm min}$.

Assuming that UVB feedback acts on a time-scale roughly corresponding to the halo sound-crossing time, we present an analytic formula for $\bar{M}_{\rm min}$.  This general expression fits our simulation results very well, and can be easily included in models of galaxy formation.

In our models, UVB feedback significantly delays reionization only if molecularly-cooled galaxies contribute significantly to reionization.  Star-formation inside such galaxies is easily suppressed by mechanical feedback, X-ray pre-heating and the baryon-DM velocity offset.  Hence, we conclude that UVB feedback is unlikely to self-regulate reionization.

\vspace{0.5cm}

We thank Andrea Ferrara for comments on a draft version of this work.

\bibliographystyle{mn2e}
\bibliography{ms}

\end{document}

%% file: defns.tex
\newcommand{\cmfast}{\textsc{\small 21CMFAST}}

\newcommand{\avenf}{\bar{x}_{\rm HI}}

\newcommand\lsim{\mathrel{\rlap{\lower4pt\hbox{\hskip1pt$\sim$}}
        \raise1pt\hbox{$<$}}}
\newcommand\gsim{\mathrel{\rlap{\lower4pt\hbox{\hskip1pt$\sim$}}
        \raise1pt\hbox{$>$}}}
\def\myputfigure#1#2#3#4#5%
{\vskip#5pt\makebox[0pt]{\hskip#2in
\includegraphics[width=#3\textwidth]{#1}}\vskip#4pt\hfill}

